\documentclass[preprint,10pt]{elsarticle}

\usepackage[utf8]{inputenc}
\usepackage[english]{babel}
\usepackage{amssymb}
\usepackage[utf8]{inputenc}
\usepackage[english]{babel}
\usepackage{natbib}
\usepackage{hyperref}
\usepackage{amsmath}
\usepackage{longtable}
\usepackage{geometry}
\usepackage{float}

\biboptions{sort&compress}
\journal{Nuclear Physics A}

\begin{document}

\begin{frontmatter}



\title{Theoretical investigation of heavy cluster decay from Z$=$118 and 120 isotopes: A search for an empirical formula in superheavy region}


\author[a]{G. Saxena}\author[b,c]{Dashty T. Akrawy}\author[b,c]{Ali H. Ahmed}\author[d]{Mamta Aggarwal}
\address[a]{Department of Physics (H\&S), Govt. Women Engineering College, Ajmer-305002, India}
\address[b]{Physics Department, College of Science, Salahaddin University, Erbil 44001, Kurdistan, Iraq}
\address[c]{Becquerel Institute for Radiation Research and Measurements, Erbil, Kurdistan, Iraq}
\address[d]{Department of Physics, University of Mumbai, Kalina, Mumbai-400098, India}

\begin{abstract}
Various decay modes in superheavy nuclei have been of significant interest among which cluster radioactivity has recently gained sizable attention. The $\alpha$-decay being a predominant decay mode in the superheavy region, the accurate determination of cluster decay half-lives is also crucial in this region as it has tremendous potential to be explored as one of the major decay channels. The usability of the Royer analytical formula [Nuclear Physics A 683 (2001) 182], which is based on the asymmetric fission model, has been investigated for the cluster and $\alpha$ decay in superheavy region, by comparing it with several other (semi)empirical/analytical formulas. After fitting the formula on around 100 cluster-decay data and around 423 $\alpha$-decay data, the refitted Royer formula (RRF) is found to be very robust which is able to estimate the cluster decay and $\alpha$-decay half-lives with good accuracy. In fact, a comparison of the half-lives of both the decay modes using the same formula points towards a substantial chance of heavy cluster (Kr and Sr) decay from various isotopes of Z$=$118 and 120. Hence, the formula proposed in this study works fairly well for the estimation of cluster decay half-lives in superheavy regions where most empirical formulas fail to match with the half-lives from the various established theories. 
\end{abstract}




\end{frontmatter}


\section{Introduction}
\label{intro}
Cluster decay known to be a heavy-particle radioactivity decay mode, has been explored and detected by several experiments over the last four decades \cite{sandulescu1980,rose1984,Bonetti2007}. The emission of heavy clusters ($^{14}$C to $^{32}$Si) from several nuclei resulting in the magic daughter nuclei (Z=82) or the neighboring nuclei close to magic number   \cite{royer2000,Royer2001,Kumar2003,Gupta2003,kulkin2005,Kumar2009,Santhosh2012,Singh2022,jain2023}, is direct evidence of the role of shell and pairing effects in the decay mechanism and, therefore, is of crucial importance in this lesser-explored region of superheavy nuclei.\par

So far, all the observations of cluster emission lie only in the trans-lead region \cite{Bonetti2007}, though its possibility in the heavy and superheavy region can not be disregarded \cite{poenaru2012,zhang2018,poenaru2018,poenaru2018EPJA,matheson2018,Warda2018,santhosh2008,prathapan2021,Santhosh2022prc,Royer2022,Nithya2022,saxena2023epja}, which is also one of the waiting events from the future experiments \cite{santhosh2018,soylu2019,jain2023PS}. The superheavy region already possesses lots of challenges in the detection system \cite{nazar2018,giuliani2019}, however, recent detection of $\alpha$-decay chains of the heaviest element $^{294}$118 \cite{ogan2006}, 11 new nuclei in the $\alpha$-decay chains of $^{293}$117 and $^{294}$117 \cite{og2010}, and the attempts of synthesizing new elements with Z$=$119 and Z$=$120 \cite{voinov2020} provide a reasonable boost for the various experimental facilities at Argonne National Laboratory in USA \cite{davids1989}, GSI in Germany \cite{hofmann2000,hofmann2011}, RIKEN in Japan \cite{morita2007}, and Dubna laboratory in Russia \cite{og2010,og2015npa}, etc. So far all the experimental facilities are mainly based on the detection of $\alpha$-particle emission, but one may expect the detection of heavy particle emission in the near future, for which the present work aims to provide useful inputs through our theoretical estimations.\par

Determination of half-lives of cluster emission can be mainly categorized into two parts. The first one is based on several theoretical models which consist of 
the models based on quantum tunneling effect through a potential barrier \cite{Poenaru1982,Delion2001,Poenaru1996,Poenaru1996IPPB,arun2009}, super-asymmetric fission model (SAFM) \cite{Poenaru1985}, unified fission model (UFM) \cite{Shi1989}, generalized liquid drop model (GLDM) \cite{Royer2001}, preformation cluster model (PCM) \cite{Poenaru2002,Joshua2022,Yahya2022}, Coulomb and proximity potential model for deformed nuclei \cite{Hosseini2023}, Wentzel-Kramers-Brilliouin (WKB) method \cite{Dagtas2022}, and modified Gamow-like model \cite{Liu2021}, etc. 
Several microscopic theories have also provided a promising way to describe cluster decay in heavy nuclei by calculating the decay width of cluster decays or the cluster preformation amplitude \cite{lovas1998,blen1988,delion1994,bhagwat2005,bbsingh2010,uzawa2022,zhao2023}. 
The second category is composed of the use of several (semi)empirical/analytical formulas that are fitted on a limited data set of cluster decay or $\alpha$-decay. Some of these formulas are Horoi formula \cite{Horoi2004}, BKAG formula \cite{Balasubramaniam2004}, RenA formula \cite{Ren2004}, NRDX formula \cite{NRDX2008}, UDL formula \cite{UDL2009}, UNIV formula \cite{Poenaru2011}, Tavares and Medeiros (TM) formula \cite{Tavares2013}, and their modified/refitted versions \cite{jain2023,soylu2021,UF2022,Wang2021CPC,cheng2022,Lin2023,saxena2023epja}. In one of our recent articles \cite{saxena2023epja}, it is shown that all the above-mentioned formulas act somewhat unfavorable in producing the half-lives of cluster emission in superheavy region with the fact that the estimated half-lives are found several orders of magnitude off than those predicted by widely used models/theories. Similarly, Ref. \cite{zhang2018} concludes that for the determination of half-lives of cluster emission in the superheavy region, only the UDL formula \cite{UDL2009} works well because of its fission-like mechanisms. This key issue has been addressed in the present article by a systematic investigation of cluster decay half-lives in superheavy nuclei.\par
 
 In the present article, we choose an analytical formula that is derived by using an asymmetric fission model \cite{royer2000,Royer2001} and hence qualifies to be applied in the superheavy region. The formula is fitted for the latest available cluster decay \cite{Bonetti2007,Gupta1994,Price1989,Santhosh2012,Hourani1989,Royer2022} and $\alpha$-decay data \cite{audi2020} and then the results are compared with the results of several other empirical formulas \cite{Horoi2004,NRDX2008,Balasubramaniam2004,Ren2004,Poenaru2011,Tavares2013,UF2022,cheng2022,soylu2021,jain2023}. The formula estimates the half-lives of the cluster as well as $\alpha$-decays in a precise manner and is found robust in the superheavy region. Several clusters are identified with great chances of decay from Og and Z=120 isotopes. 
\section{Formalism and Calculations}
The generalized liquid drop model (GLDM) is used to describe nuclear fusion, fission, cluster decay, $\alpha$-decay, and proton emission processes with the inclusion of nuclear proximity energy and quasi molecular shape \cite{royer1984,royer2000,Royer2001}. The macroscopic total energy is determined as:
\begin{eqnarray}
 E=E_{V}+E_{S}+E_{C}+E_{N}
 \label{energy}
 \end{eqnarray}
 where, for one-body shapes, the volume (E$_{V}$), surface (E$_{S}$), and Coulomb (E$_{C}$) energies are given by:
 \begin{eqnarray}
 E_{V}=-15.494(1-1.8I^2)A MeV
 \label{Ev}
 \end{eqnarray}
 \begin{eqnarray}
 E_{S}=17.9439(1-2.6I^2)A^{2/3}(S/4 \pi R^{2}_{o}) MeV
 \label{Es}
 \end{eqnarray}
\begin{eqnarray}
 E_{C}=0.6e^{2}(Z^{2}/R_{0})\times 0.5 \int (V(\theta)/V_{0})(R(\theta)/R_{0})^{3} sin \theta d \theta 
 \label{Ec}
 \end{eqnarray}
 in the above formulas, $I$ and $S$ represent relative neutron excess and the surface of the deformed nucleus, respectively. Additionally, $V(\theta)$ denotes the electrostatic potential at the surface. Whereas, $V_{0}$ is the surface potential of the sphere and $R_{0}$ is the effective sharp radius which is chosen as:
 \begin{eqnarray}
 R_{o}=1.28A^{1/3}-0.76+0.8A^{-1/3} fm
 \label{Ro}
 \end{eqnarray}
In cluster emission, the parent nucleus breaks mainly into two fragments and when the fragments are separated the the energies can be written as:
\begin{eqnarray}
 E_{V}=-15.494[(1-1.8I^{2}_{1})A_{1}+(1-1.8I^{2}_{2})A_{2}] MeV
 \label{E.v}
 \end{eqnarray}
 \begin{eqnarray}
 E_{S}=17.9439[(1-2.6I^{2}_{1})A^{2/3}_{1}+(1-2.6I^{2}_{2})A^{2/3}_{2}] MeV
 \label{E.s}
 \end{eqnarray}
\begin{eqnarray}
 E_{C}=0.6e^{2}Z^{2}_{1}/R_{1}+0.6e^{2}Z^{2}_{2}/R_{2}+e^{2}Z_{1}Z_{2}/r
 \label{E.c}
 \end{eqnarray}
 where A$_{i}$, Z$_{i}$, R$_{i}$, and I$_{i}$ are the masses, charges, radii, and relative neutron excesses of the fragments, and r is the distance between the mass centers. The radii of the two emitted fragments have been calculated from the radius of the decaying nucleus to ensure volume conservation and final asymmetry:
 \begin{eqnarray}
 R_{1}=R_{0}(1+\beta ^{3})^{-1/3}
 \label{R1}
 \end{eqnarray}
 \begin{eqnarray}
 R_{2}=R_{0}\beta(1+\beta ^{3})^{-1/3}
 \label{R2}
 \end{eqnarray}
 where
 \begin{eqnarray}
 \beta = \frac{1.28A^{1/3}_{2}-0.76+0.8A^{-1/3}_{2}}{1.28A^{1/3}_{1}-0.76+0.8A^{-1/3}_{1}}
 \label{B}
 \end{eqnarray}
 The discontinuity of a few MeV appearing at the contact point between the nascent spherical fragments when Z$_{1}$/A$_{1}$ and Z$_{2}$/A$_{2}$ are very different has been subtracted linearly from the contact point to the sphere to follow the progressive rearrangement of the nuclear matter before separation. The surface energy $E_S$ takes into account the effects of surface tension forces in a half-space only and does not include the contribution of the attractive nuclear forces between the considered surfaces in the neck or in the gap between the fragments. The term nuclear proximity energy $E_N$ allows us to consider these additional surface effects when a neck or a gap appears:
 \begin{eqnarray}
 E_{N}(r) = 2 \gamma \int_{h_{min}}^{h_{max}} \Phi [D(r,h)/b]2 \pi h dh
 \label{r} 
 \end{eqnarray}
 where $h$ is the transverse distance varying from the neck radius or zero to the height of the neck border, $D$ is the distance between the opposite infinitesimal surfaces considered, and b is the surface width fixed at 0.99 fm. $\Phi$ is the proximity function of Feldmeier. The surface parameter $\gamma$ is the geometric mean between the surface parameters of the two fragments:
 \begin{eqnarray}
 \gamma=0.9517 \sqrt{(1-2.6I^{2}_{1})(1-2.6I^{2}_{2})} MeV fm^{-2}
 \label{gamma}
 \end{eqnarray}

Based on the above asymmetric fission model and fragments within the quasimolecular shape sequence, a simple analytical formula to determine the half-life of the light-nucleus emission was proposed \cite{poenaru1996,blendowske1996}, which was subsequently fitted on a set of 144 cluster decay half-lives predicted by the GLDM \cite{royer2000,Royer2001} with a rms deviation of 0.42 and given as the following formula:
\begin{equation}
 Log_{10}T_{1/2}(s) = a + R_{cont}\sqrt{A_{12}Q} \left[ b + \frac{c}{\sqrt{Q}}+ d \sqrt{\frac{Q}{A_{12}R_{cont}}} + \frac{e}{\sqrt{x}} + f\sqrt{x} + gx     \right]
 \label{RRF}
\end{equation}
where, 
\begin{equation}
x=  \frac{R_{cont}Q}{e^{2}Z_{C}Z_{D}} 
\end{equation}
and
\begin{equation}
A_{12}= \frac{A_{1}A_{2}}{A_{1}+A_{2}}    
\end{equation}
with $Z_{C}$ and $Z_{D}$ are the atomic number of cluster and daughter nucleus and $A_{12}$ is reduced mass. $Q$ is the disintegration energy of the system. $R_{cont}$ is the distance between the centers of the cluster and daughter nuclei at the contact point which is given as $R_{cont} = R_{1} +R_{2}$, where $R_{1}$ and $R_{2}$ are the radii of the two emitted fragments calculated by using Eqns \ref{R1} and \ref{R2}. It is to be pointed out here that this formula \ref{RRF} has been utilized to calculate cluster decay half-lives \cite{Royer2001} in the trans-lead region.\par

For cluster decay, there are only 37 parent-cluster combinations data from various experiments \cite{Bonetti2007,Gupta1994,Price1989,Santhosh2012,Hourani1989} which lead to a total of only 61 data points due to multiple $Q$-values corresponding to various detection systems. These all data points merely belong to the trans-lead region in the range 87$\leq$Z$\leq$96, and therefore are insufficient to fit any formula aiming for the superheavy region. To overcome this situation, in addition to the 61 experimental data for fitting, we have included half-lives of 54 parent-cluster combinations ranging 104$\leq$Z$\leq$118, recently reported in Ref. \cite{Royer2022} using GLDM. This selection of data is justified by the fact that GLDM data has already been used to fit the cluster decay formula in the absence of experimental data in Ref. \cite{Royer2001}, also the GLDM data is found very close to the experimental data in the recent Ref. \cite{scientifi-report}. Additionally, the GLDM approach can be treated as a reliable treatment because it can provide insights into the sensitivity of cluster emission half-lives to various factors such as the size and composition of the emitted cluster, as well as the characteristics of the parent nucleus. This model also takes into account surface effects providing a more realistic description of the potential energy landscape for cluster emission. Due to the above advantages, GLDM has allowed the reproducing of some of the fusion \cite{royer1985}, fission \cite{royer2012}, cluster \cite{Royer2001,royer1998}, and $\alpha$-radioactivity \cite{royer2000,deng2020} data.\par

After fitting on 61 (experimental) + 54 (GLDM) data, the corresponding coefficients for the refitted Royer formula (RRF) are mentioned in Table \ref{tab:coefficient} for the cluster decay. For the $\alpha$-decay, since adequate data are available, therefore, we have fitted the same formula for 423 experimental data taken from NUBASE2020 \cite{audi2020} and the coefficients are mentioned in Table \ref{tab:coefficient}. Since, for superheavy nuclei, the uncertainties in the estimation play a crucial role, therefore, these coefficients are calculated by taking experimental uncertainties \cite{audi2020} into consideration which are found to be with the value of standard deviation $\pm$0.4 Sec. in the logarithm of half-lives. 

\begin{table}[!htbp]
 \caption{The coefficients of the RRF formula proposed in the present work. The formula is fitted separately for Cluster and $\alpha$-decay for a total of 61+54 data \cite{Bonetti2007,Gupta1994,Price1989,Santhosh2012,Hourani1989,Royer2022} and 423 data \cite{audi2020}, respectively (see the text for details).}
 \centering
  \resizebox{0.6\textwidth}{!}{%
 \begin{tabular}{@{\hskip 0.008in}|c|@{\hskip 0.5in}c|@{\hskip 0.5in}c|}
 \hline
Coefficient&Cluster decay&$\alpha$-decay\\
\cline{2-3}
   \hline
 a&21.4346$\pm$0.3996    & -27.0901$\pm$0.3999      \\
 b&-0.3740$\mp$0.0001    & -8.8888$\pm$1.45E-05    \\ 
 c&-2.7167$\pm$1.9E-05    & 0.6581$\pm$1.74E-05        \\ 
 d&-0.1028$\pm$1.6E-06 & 0.3498$\pm$8.73E-06      \\ 
 e&0.5827$\pm$2.6E-05  & 2.2470$\mp$4.10E-06        \\ 
 f&0.0540$\pm$0.0001 & 12.6087$\mp$3.80E-05          \\
 g&-0.1146$\mp$5.1E-05 & -6.8422$\pm$2.15E-05          \\
\hline
 \end{tabular}}
\label{tab:coefficient}
\end{table}

\section{Results and discussions}
The applicability of the proposed formula RRF can be asserted on the grounds of comparison with other known and widely used formulas. In Table \ref{rmse-cluster}, we list values of root-mean-square deviation ($\sigma$) and uncertainty ($u$) for the several formulas viz. Royer \cite{Royer2001}, UDL \cite{UDL2009}, TM \cite{Tavares2013}, ITM \cite{saxena2023epja}, MBKAG \cite{jain2023}, BKAG \cite{Balasubramaniam2004}, MUDL \cite{jain2023}, RenA \cite{Ren2004}, MRenA \cite{jain2023}, Soylu \cite{soylu2021}, ISEF \cite{cheng2022}, NRDX \cite{NRDX2008}, MNRDX \cite{jain2023}, IUF \cite{UF2022}, UNIV \cite{Poenaru2011}, MHoroi \cite{jain2023}, Horoi \cite{Horoi2004} along with RRF (present work). The $\sigma$ and $u$ are calculated by using the following equations:\\

\begin{equation}
\sigma  = \sqrt{\frac{1}{N_{nucl}}\sum^{N_{nucl}}_{i=1}\left(log\frac{T^i_{Th.}}{T^i_{Exp.}}\right)^2}
\end{equation}
\begin{equation}
u =  \sqrt{\frac{1}{N_{nucl}(N_{nucl}-1)}\sum^{N_{nucl}}_{i=1}\left(log\frac{T^i_{Th.}}{T^i_{Exp.}}-\mu \right)^2}
\end{equation}
where $N_{nucl}$ is the total number of nuclei (data). $T^i_{Exp.}$ and $T^i_{Th.}$ are the experimental and theoretical values of half-lives for $i^{th}$ data point, respectively. The least values of $\sigma$ and uncertainty set up the RRF for precise estimation of cluster decay half-lives. Additionally, the RRF is also found to provide the least values of $\sigma$ and uncertainty for the $\alpha$-decay half-live as can be seen in the half-right part of Table \ref{rmse-cluster}. Hence, this comparison with known data (cluster and $\alpha$) proves the efficiency of the RRF formula and demonstrates its ability to accurately estimate half-lives in the unknown region of the periodic chart.\par 

\begin{table}[!htbp]
\caption{Comparison of RRF formula with several other formulas for 115 cluster decay data \cite{Bonetti2007,Gupta1994,Price1989,Santhosh2012,Hourani1989,Royer2022} and 415 $\alpha$-decay data \cite{audi2020} (see the text for details).}
 \centering
  \resizebox{0.55\textwidth}{!}{%
 \begin{tabular}{|ccc|ccc|}
 \hline
\multicolumn{3}{|c|}{Cluster Decay}&
\multicolumn{3}{c|}{$\alpha$-Decay}\\
\hline
Formula    &$\sigma$& $u$& Formula    &$\sigma$& $u$\\
\hline
RRF (Present work)    &  0.88      & $\pm$0.09 &        RRF (Present work)	         &       0.61      &        $\pm$0.03       \\ 
Royer \cite{Royer2001}   &  1.42      & $\pm$0.14 &     UDL \cite{UDL2009}  	     &         0.74      &        $\pm$0.03       \\ 
UDL \cite{UDL2009}     &  2.36      & $\pm$0.24 &       QF \cite{saxena2021}  	        &       0.75      &        $\pm$0.04       \\ 
TM \cite{Tavares2013}      &  2.55      & $\pm$0.27 &   UNIV \cite{Poenaru2011}	        &       0.80      &        $\pm$0.03       \\ 
ITM \cite{saxena2023epja}    &  3.88      & $\pm$0.40 &  Horoi \cite{Horoi2004}	     &         0.84      &        $\pm$0.04       \\ 
MBKAG \cite{jain2023}   &  4.63      & $\pm$0.49 &      NRDX \cite{NRDX2008}	         &       0.88      &        $\pm$0.04       \\ 
BKAG \cite{Balasubramaniam2004}   &  4.58 & $\pm$0.47 & MRenB2019 \cite{newrenA2019}     &       0.88      &        $\pm$0.04       \\ 
MUDl \cite{jain2023}   &  5.29      & $\pm$0.47 &       RB \cite{Royer2010}	           &       0.91      &        $\pm$0.04       \\ 
RenA  \cite{Ren2004}   &  5.17      & $\pm$0.45 &       DK1	\cite{akrawy2018}          &       0.92      &        $\pm$0.04       \\ 
MRenA \cite{jain2023}  &  6.64      & $\pm$0.59 &       NMSF \cite{pksharma2021}	    &       0.94      &        $\pm$0.04       \\ 
Soylu \cite{soylu2021}   &  7.54      & $\pm$0.69 &     MRF \cite{Akrawymrf2018}    &       1.01      &        $\pm$0.04       \\ 
ISEF \cite{cheng2022}    &  11.73     & $\pm$1.02 &     Soylu \cite{soylu2021}	        &       1.07      &        $\pm$0.04       \\ 
NRDX \cite{NRDX2008}   &  11.82     & $\pm$1.01 &       NMMF \cite{pksharma2021}     &       1.18      &        $\pm$0.06       \\ 
MNRDX \cite{jain2023}   &  11.92     & $\pm$1.03 &      Royer \cite{Royer2001}     &       1.47      &        $\pm$0.04       \\ 
IUF \cite{UF2022}     &  12.79     & $\pm$1.10 &        AAF 2018 \cite{akrawy2018ijmpe}   &       1.76      &        $\pm$0.08       \\ 
UNIV \cite{Poenaru2011}   &  13.16     & $\pm$1.15 &    YQZR \cite{YQZR}	         &       2.93      &        $\pm$0.13       \\ 
Mhoroi \cite{jain2023}  &  16.31     & $\pm$1.42 &      M.Budaca \cite{Akrawy2022}      &       4.26      &        $\pm$0.19       \\ 
Horoi \cite{Horoi2004}   &  17.95     & $\pm$1.56 &                      RenA \cite{Ren2004}         &       6.77      &        $\pm$0.29       \\ 
\hline
\end{tabular}}
\label{rmse-cluster}
\end{table}

Even with good accuracy of the RRF, the estimation of half-lives of cluster decay in the unknown region of the periodic chart also relies on the $Q$-values as can be seen from Eqn, \ref{RRF}. Therefore, picking up a model or theoretical calculation that leads to precise $Q$-values is requisite for determining the unknown half-lives with good accuracy. In view of this, we have calculated $Q$-values by using the following formula by taking binding energies (for daughter(d), cluster(c), and parent(p) nuclei) from various theoretical treatments/mass tables. 
\begin{eqnarray}
 Q (MeV) = B.E. (d)+ B.E. (c)-B.E. (p) + k[Z_{p}^{\epsilon}-Z_{d}^{\epsilon}]
 \label{Q_value}
 \end{eqnarray}
where, the term $k[Z_{p}^{\epsilon}-Z_{d}^{\epsilon}]$ indicates the screening effect caused by the surrounding electrons around the nuclei \cite{Denisov2009prc} with k=8.7 eV [8.7 $\times$ 10$^{-6}$MeV] and $\epsilon$=2.517 for Z (proton number) $\geq$ 60, and k=13.6 eV [13.6 $\times$
10$^{-6}$MeV] and $\epsilon$ =2.408 for Z $<$ 60 have been deducted from the data shown by Huang \textit{et al.} \cite{Huang1976}. We have selected WS4\cite{WS4}, FRDM \cite{FRDM2016}, INM \cite{INM2012}, HFB \cite{hfb}, DD-ME2 \cite{Agbemava2014}, NL3$^{\*}$ \cite{Agbemava2014}, KTUY \cite{KTUY2005},  UNEDF1 \cite{SKM,UNEDF1}, NSM \cite{Aggarwal2010,NSM}, RMF \cite{Singh2012,saxena2019,saxena2019plb}, DD-PC1 \cite{Agbemava2014}, DD-MEdelta \cite{Agbemava2014}, SV-Min \cite{SKM,SV}, UNEDF0 \cite{SKM,UNEDF0}, SKP \cite{SKM,SKP}, RCHB \cite{rchb}, SLY4 \cite{SKM,SLY4}, and SKM$^{\*}$ \cite{SKM,SKMS} mass models. For these approaches, we have calculated $\sigma$ and $u$ for the known 61 $Q$-values \cite{Bonetti2007,Price1989,Royer2001,soylu2021} as listed in Table \ref{T3}. It is clear from Table \ref{T3} that the WS4 mass model provides an excellent agreement with the minimum $\sigma$ as compared to all other considered theoretical approaches and hence justifies the calculation of $Q$-values for cluster emission \cite{WS4}. Henceforth, we will use unknown $Q$-values from the WS4 mass model for the estimation of half-lives using the RRF. 


\begin{table}[!htbp]
\caption{RMSE and uncertainity ($u$) of various mass models for 61 $Q$-value data for cluster emission.}
\centering
\resizebox{0.76\textwidth}{!}{%
\begin{tabular}{l@{\hskip 2.0in}c@{\hskip 2.0in}c}
\hline
\hline
Theory&$\sigma$&$u$\\
\hline
WS4 \cite{WS4}&0.34&$\pm$0.04\\
FRDM \cite{FRDM2016}&0.76&$\pm$0.10\\
INM \cite{INM2012}&0.85&$\pm$0.10\\
HFB \cite{hfb}&1.19&$\pm$0.09\\
DD-ME2 \cite{Agbemava2014}&1.77&$\pm$0.30\\
KTUY \cite{KTUY2005}&1.99&$\pm$0.17\\
NL3$^{\*}$ \cite{Agbemava2014}&2.11&$\pm$0.39\\
UNEDF1 \cite{SKM,UNEDF1}&2.15&$\pm$0.13\\
NSM \cite{Aggarwal2010,NSM}&2.55&$\pm$0.11\\
RMF \cite{Singh2012,saxena2019,saxena2019plb}&3.06&$\pm$0.26\\
DD-PC1 \cite{Agbemava2014}&3.32&$\pm$0.30\\
DD-MEdelta \cite{Agbemava2014}&4.09&$\pm$0.33\\
SV-Min \cite{SKM,SV}&5.84&$\pm$0.12\\
UNEDF0 \cite{SKM,UNEDF0}&6.99&$\pm$0.12\\
SKP \cite{SKM,SKP}&8.17&$\pm$0.16\\
RCHB \cite{rchb}&9.11&$\pm$0.46\\
SLY4 \cite{SKM,SLY4}&11.32&$\pm$0.26\\
SKM$^{\*}$ \cite{SKM,SKMS}&13.78&$\pm$0.17\\
\hline
\hline
\end{tabular}}
\label{T3}
\end{table} 

For more precise outcomes, it is important to further examine/test the already calculated half-lives of cluster decay in the superheavy region by using various established theories such as the estimation of half-lives using the double-folding model with the M3Y-Paris NN force assuming the finite-range exchange part with spherical density distributions (Sph-FR), the zero-range exchange contribution with either spherical (Sph-ZR) or deformed (Def-ZR) density distributions of the daughter nuclei \cite{Ismail2019} as well as CPPM \cite{santhosh2018} and WKB$^{*}$ \cite{soylu2019}. The test can be of more worth if the RRF half-lives are compared with the half-lives estimated by using widely used empirical formulas viz. UDL formula \cite{UDL2009}, NRDX formula \cite{NRDX2008}, Horoi formula \cite{Horoi2004}. It is important to point out here that among several empirical/analytical formulas, only the UDL formula is able to produce a competitive relationship between $\alpha$-decay and cluster radioactivity in the superheavy region, due to its treatment as both the preformation model and the fission-like mechanisms \cite{Zhang2018}. In Table \ref{table-comparison}, we have compared the logarithmic half-lives of cluster decay calculated by using RRF for the various isotopes of superheavy nuclei Z=118 (Og) and Z=120. In addition to the Horoi formula \cite{Horoi2004} and NRDX formula \cite{NRDX2008} as mentioned in Table \ref{table-comparison}, we have also checked the other existing formulas viz.  BKAG formula \cite{Balasubramaniam2004}, RenA formula \cite{Ren2004}, UNIV formula \cite{Poenaru2011}, Tavares and Medeiros formula \cite{Tavares2013}, IUF \cite{UF2022}, ISEF \cite{cheng2022}, new UDL formula \cite{soylu2021}, and our modified formulas (MUDL, MRenA, MUNIV, MHoroi, MNRX) \cite{jain2023} but none of the formulas is found to work in this region i.e. the cluster decay half-lives are quite far and several orders of magnitude off from what is reported by several theories mentioned above. \par
\begin{figure}[!htbp]
\centering
\includegraphics[width=0.6\textwidth]{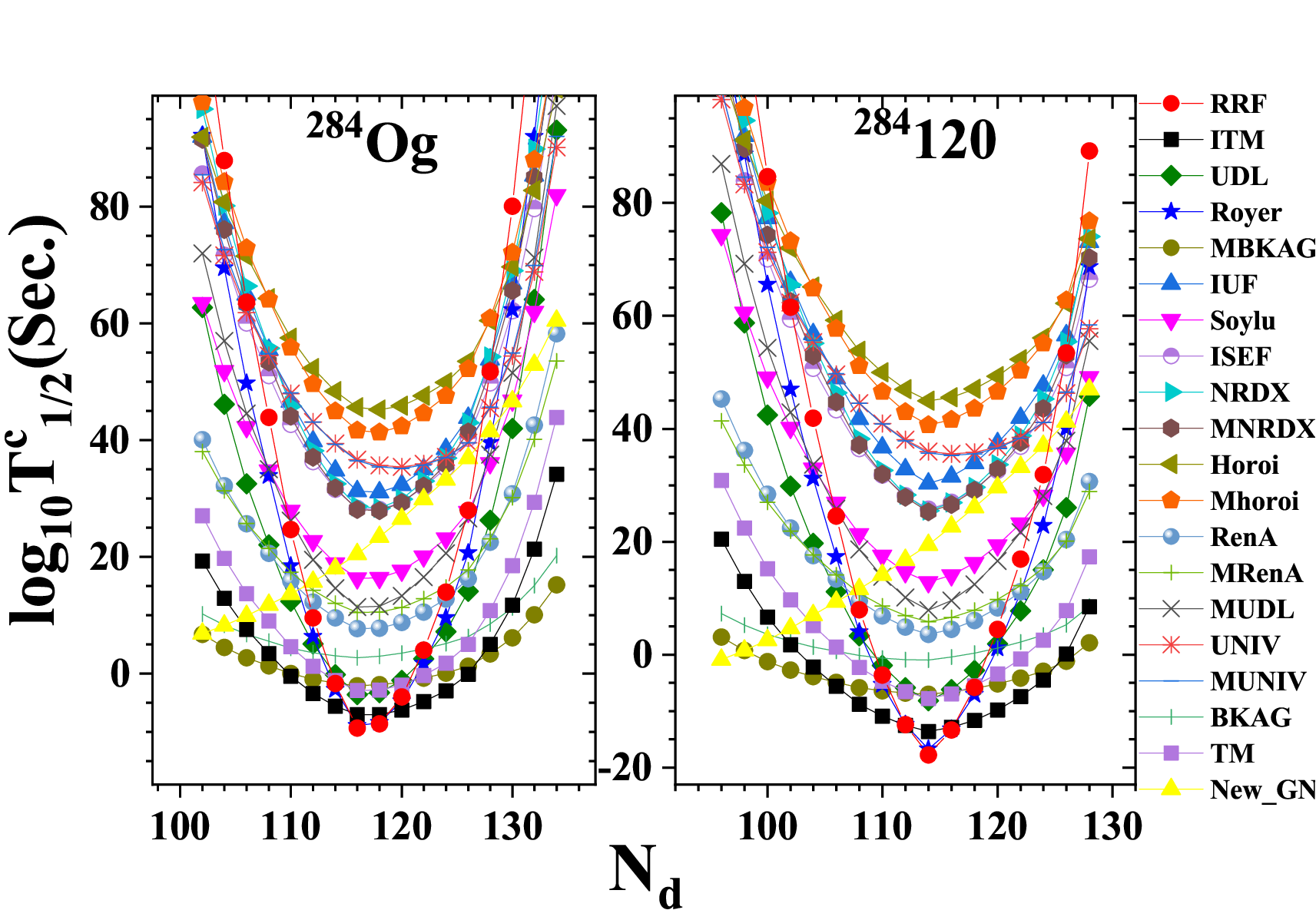}
\caption{(Colour online) Cluster decay half-lives calculated by using various empirical/analytical formulas i.e. ITM \cite{saxena2023epja}, UDL \cite{UDL2009}, Royer \cite{Royer2001}, MBKAG \cite{jain2023}, IUF \cite{UF2022}, Soylu \cite{soylu2021}, ISEF \cite{cheng2022}, NRDX \cite{NRDX2008}, MNRDX \cite{jain2023}, Horoi \cite{Horoi2004}, MHoroi \cite{jain2023}, RenA \cite{Ren2004}, MRenA \cite{jain2023}, MUDL \cite{jain2023}, UNIV \cite{Poenaru2011}, MUNIV \cite{jain2023}, BKAG \cite{Balasubramaniam2004}, TM \cite{Tavares2013}, and New GN \cite{Lin2023} for isotopes of Kr and Se emitting from $^{284}$Og and $^{284}120$, respectively, leading to daughter nuclei of Pb isotopes. (N$_d$) represents the neutron number of the daughter nucleus.}
\label{half-lives}
\end{figure}

\begin{table*}[!htbp]
\caption{The calculated logarithmic half-lives using RRF formula (RRF) compared with UDL formula \cite{UDL2009}, NRDX formula \cite{NRDX2008}, Horoi formula \cite{Horoi2004}  and various theories viz. CPPM \cite{santhosh2018}, WKB$^{*}$ \cite{soylu2019}, and double-folding model with the M3Y-Paris NN force assuming the finite-range exchange part with spherical density distributions (Sph-FR), the zero-range exchange contribution with either spherical (Sph-ZR) or deformed (Def-ZR) density distributions of the daughter nuclei \cite{Ismail2019}.}
\centering
\resizebox{0.9\textwidth}{!}{%
\begin{tabular}{cccccccccccccc}
\hline
\multicolumn{1}{c}{Parent}&
\multicolumn{1}{c}{Daughter}&
\multicolumn{1}{c}{Emitted}&
\multicolumn{1}{c}{$Q$}&
\multicolumn{1}{c}{$l$}&
\multicolumn{9}{c}{log$_{10}$T$_{1/2}$(Sec.)}\\
\cline{6-14}
\multicolumn{1}{c}{nucleus}&
\multicolumn{1}{c}{nucleus}&
\multicolumn{1}{c}{cluster}&
\multicolumn{1}{c}{(MeV)}&
\multicolumn{1}{c}{}&
\multicolumn{1}{c}{RRF}&
\multicolumn{1}{c}{UDL}&
\multicolumn{1}{c}{NRDX}&
\multicolumn{1}{c}{Horoi}&
\multicolumn{1}{c}{Sph-FR}&
\multicolumn{1}{c}{Sph-ZR}&
\multicolumn{1}{c}{Def-ZR}&

\multicolumn{1}{c}{CPPM}&
\multicolumn{1}{c}{WKB$^{*}$}\\

 \hline
$^{302}$Og   &  $^{208}$Pb & $^{94}$Kr & 294.48 & 0  & -1.67 & 0.44   &       &  &   &     &  &3.82  & 1.28         \\      
$^{304}$Og   &  $^{208}$Pb & $^{96}$Kr & 293.32 & 0  & -0.82 & 1.41   &       &  &   &     &  &4.65  & 2.55         \\      
$^{306}$Og   &  $^{208}$Pb & $^{98}$Kr & 290.38 & 0  &  3.71 & 3.81   &       &  &   &     &  &6.83  & 5.16         \\       
$^{308}$Og   &  $^{208}$Pb &$^{100}$Kr & 286.03 & 0  & 11.27 & 7.40   &       &  &   &     &  &10.00 & 8.88         \\
$^{295}$119   &  $^{279}$Rg & $^{16}$O  & 59.26  & 0  & 29.39&24.62 &27.19& 23.46 &24.14& 26.45& 25.78 & &         \\
$^{295}$119   &  $^{277}$Rg & $^{18}$O  & 58.90  & 2  & 30.29&27.32 &30.45& 27.01 &26.64& 29.06& 28.03 & &         \\
$^{295}$119   &  $^{275}$Rg & $^{20}$O  & 56.97  & 0  & 33.86&32.58 &36.22& 32.82 &31.61& 34.17& 32.89 & &         \\
$^{295}$119   &  $^{273}$Mt & $^{22}$Ne & 79.79  & 0  & 30.95&26.00 &30.62& 27.55 &26.51& 29.41& 27.84 & &         \\
$^{295}$119   &  $^{271}$Mt & $^{24}$Ne & 80.57  & 0  & 29.76&26.30 &31.61& 28.90 &26.71& 29.68& 27.98 & &         \\
$^{295}$119   &  $^{269}$Mt & $^{26}$Ne & 75.98  & 0  & 38.06&34.50 &40.32& 36.94 &34.43& 37.58& 35.74 & &         \\
$^{295}$119   &  $^{267}$Bh & $^{28}$Mg & 101.80 & 0  & 28.89&24.16 &31.25& 29.41 &26.07& 29.45& 27.27 & &         \\
$^{295}$119   &  $^{265}$Bh & $^{30}$Mg & 98.00  & 0  & 35.63&30.07 &37.75& 35.26 &31.52& 35.07& 32.67 & &         \\
$^{295}$119   &  $^{263}$Db & $^{32}$Si & 122.34 & 0  & 28.00&22.46 &31.46& 30.72 &25.95& 29.73& 27.12 & &         \\
$^{295}$119   &  $^{261}$Db & $^{34}$Si & 121.00 & 0  & 30.29&24.69 &34.41& 33.51 &27.96& 31.83& —     & &         \\
$^{298}$120   &  $^{282}$Cn & $^{16}$O  &  60.142 & 0&  28.95 &24.18&25.66 &23.19& 22.48 &24.79& 24.14&  &         \\
$^{298}$120   &  $^{280}$Cn & $^{18}$O  &  59.622 & 2&  30.09 &27.08&29.13 &26.92& 25.18 &27.61& 26.88&  &         \\
$^{298}$120   &  $^{278}$Cn & $^{20}$O  &  57.402 & 0&  34.18 &32.81&35.38 &33.14& 30.61 &33.19& 32.13&  &         \\
$^{298}$120   &  $^{278}$Ds & $^{20}$Ne &  74.898 & 0&  39.75 &32.22&34.95 &31.63& 31.35 &34.30& 33.06&  &         \\
$^{298}$120   &  $^{276}$Ds & $^{22}$Ne &  80.005 & 0&  31.90 &26.74&30.30 &28.34& 26.00 &28.93& 27.61&  &         \\
$^{298}$120   &  $^{274}$Ds & $^{24}$Ne &  81.345 & 0&  29.78 &26.27&30.56 &29.07& 25.49 &28.48& 26.89&  &         \\
$^{298}$120   &  $^{272}$Ds & $^{26}$Ne &  77.559 & 0&  36.63 &33.18&38.01 &36.00& 31.97 &35.12& 33.33&  &         \\
$^{298}$120   &  $^{270}$Hs & $^{28}$Mg & 103.683 & 0&  27.31 &23.00&29.13 &28.72& 23.82 &27.21& 25.20&  &         \\
$^{298}$120   &  $^{268}$Hs & $^{30}$Mg &  99.651 & 0&  34.40 &29.08&35.83 &34.71& 29.44 &32.99& 30.80&  &         \\
$^{298}$120   &  $^{266}$Sg & $^{32}$Si & 124.204 & 0&  26.76 &21.62&29.70 &30.36& 24.03 &27.82& 25.31&  &         \\
$^{310}$120  &  $^{208}$Pb &$^{102}$Sr & 312.61 & 0  & -7.53 & 2.25   &       &  &   &     &  &3.27  & 4.27         \\ 
$^{312}$120  &  $^{208}$Pb &$^{104}$Sr & 308.64 & 0  & -0.91 & 0.32   &       &  &   &     &  &1.36  & 2.09         \\      
$^{313}$120  &  $^{208}$Pb &$^{105}$Sr & 306.29 & 2  &  3.20 & 2.25   &       &  &   &     &  &3.27  & 4.29         \\      
\hline
\end{tabular}}
\label{table-comparison}
\end{table*}

\begin{figure}[h]
\centering
\includegraphics[width=1.00\textwidth]{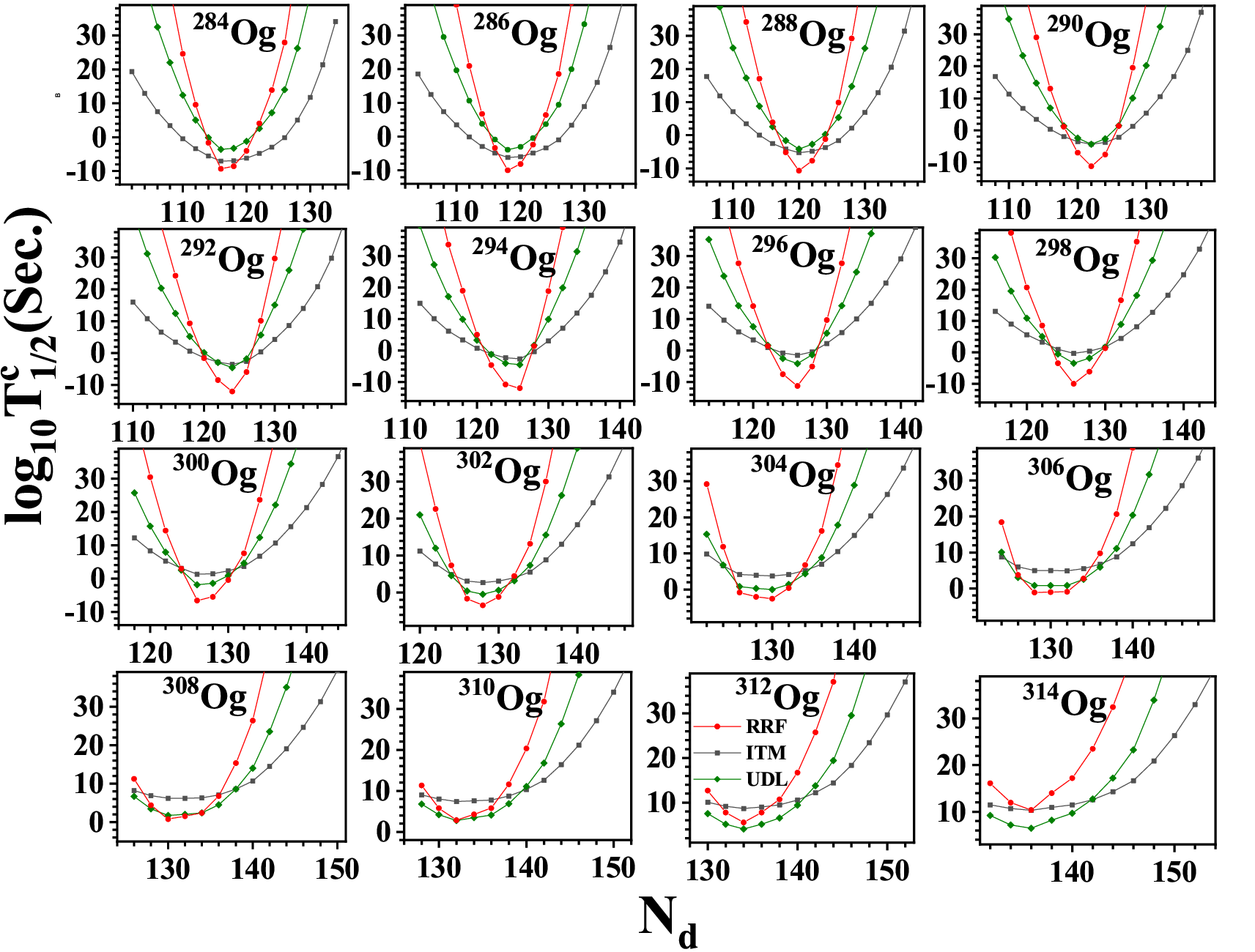}
\caption{(Colour online) Cluster decay half-lives for various even-even isotopes (N$_{p}$=166-196) of Og calculated by the RRF formula and comparison with UDL formula.}
\label{half-lives-118}
\end{figure}

Table \ref{table-comparison} endorses the estimation by RRF in the superheavy region as the half-lives of heavy cluster decay from Z$=$118 and 120 isotopes are found in close proximity with those predicted by the listed theories. For more insight, we have plotted half-lives of heavy clusters i.e isotopes of Kr and Se emitting from $^{284}$Og and $^{284}120$, respectively, leading to daughter nuclei of Pb isotopes in Fig. \ref{half-lives}.

X-axis represents the neutron number of the daughter fragment which corresponds to a single isotope of the cluster. The half-lives are calculated using several known formulas for cluster decay using $Q$-values from the WS4 mass model, as mentioned above. A clear parabolic trend is seen in almost all the empirical formulas which indicate greater chances of cluster decay having daughter neutron number N$_d$$\sim$120. However, the considerable variation in half-lives log$_{10}T_{1/2}$ from $\sim$ -20 to 60, suggests a clear dissimilarity in the empirical formulas, especially for estimating cluster decay half-lives in the superheavy region. On the other side, a few formulas viz. UDL, ITM, MBKAG, and the present formula RRF result in the half-lives in a similar order obtained by several theories (as listed in Table \ref{table-comparison}).

Furthermore, the values of these half-lives, corresponding to the minimum of the parabola, are also in the similar order of $\alpha$-decay as predicted by many other theories/formulas \cite{saxena2021,saxena2023epja,Akrawymrf2018,akrawy2018,newrenA2019,IRF2022,MYQZR2019,royer2020,akrawy2018ijmpe,UF2022,Ismail2022,cheng2022,Santhosh2022,Azeez2022,Ghorbani2022,YAHYA2022Prmana,pksharma2021,Royer2010,YQZR,Akrawy2022}. Hence, for the systematic investigation of cluster radioactivity in superheavy region, we can rely on ITM, UDL formula along with the proposed RRF formula. In view of this, in Fig. \ref{half-lives-118}, we have displayed cluster decay half-lives for various even-even isotopes (N$_{p}$=166-196) of Og (Z$=$118) corresponding to the daughter nucleus as one of the Pb isotopes, which consequently leads to the decay of several Kr isotopes. The trend of half-lives with all the formulas is found in a parabolic shape suggesting a greater probability of at least one cluster for each considered isotope of Og. Importantly, all the considered formulas somewhat provide the same minimum for all the isotopes and render one cluster for each isotope with the greatest probability of decay. In addition, the range of the half-lives is found similar to the $\alpha$-decay half-lives of superheavy nuclei and also within the experimental limit. \par
\begin{figure}[h]
\centering
\includegraphics[width=1.0\textwidth]{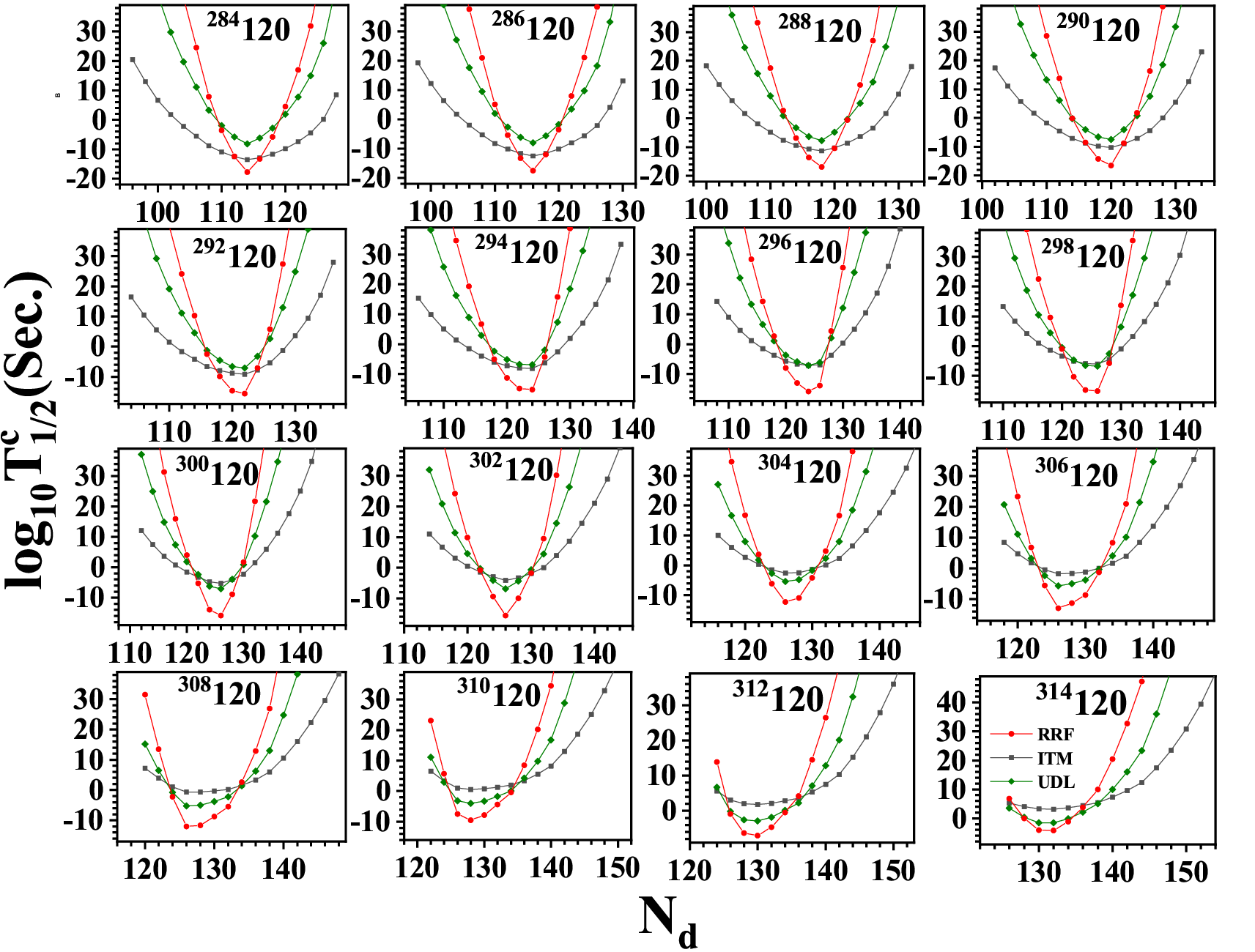}
\caption{(Colour online) Cluster decay half-lives for various even-even isotopes (N$_{p}$=164-194) of Z=120 calculated by the RRF formula and comparison with UDL formula.}
\label{half-lives-120}
\end{figure}

\begin{figure}[!htbp]
\centering
\includegraphics[width=0.70\textwidth]{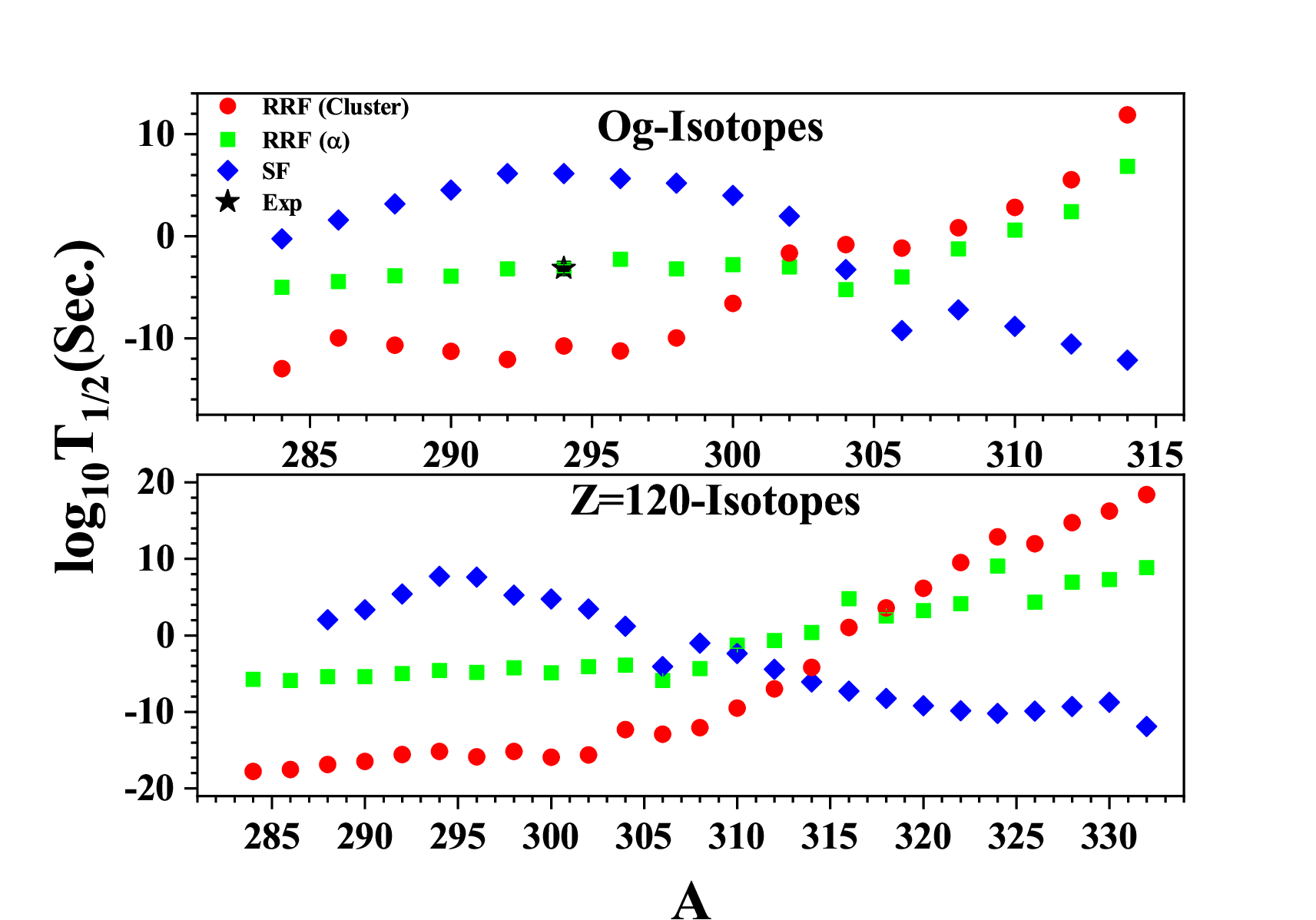}
\caption{(Colour online) Comparison of Cluster decay half-lives of RRF formula with the half-lives of $\alpha$-decay and spontaneous fission for Og and Z=120 isotopes.}
\label{half-lives-all}
\end{figure}

In an analogous manner, we have calculated half-lives for Z=120 isotopes (N$_{p}$=164-194) and displayed in Fig. \ref{half-lives-120}. Again a similar parabolic trend and the minimum lead to the most probable cluster combination for each of the considered isotopes of Z=120. With the minimum of the RRF curve in each panel of Figs. \ref{half-lives-118} and \ref{half-lives-120}, one can find the most likely cluster and its corresponding parent(daughter) nucleus of Og and Z=120 isotopes, respectively. However, only this selection of cluster does not guarantee its emission because in this region of periodic chart $\alpha$-decay and spontaneous fission dominants in all the possible decay modes. Hence, a real test of cluster emission is when the half-lives are compared with the other decay modes. In this regard, we have an advantage in that we have the same RRF for the estimation of half-lives of $\alpha$-decay, hence, the comparison of cluster and $\alpha$ decays is governed by the same kind of Physics based on asymmetric fission. For the spontaneous fission (SF), we use the recently reported modified Bao formula (MBF) \cite{Saxena2021jpg} and plot all these half-lives in Fig. \ref{half-lives-all}.

The probability of SF in both the isotopic chains is found very little for the lower mass region (A$<$304) and therefore in this region cluster decay and $\alpha$-decay compete with each other. For nuclei with A$>$306 and A$>$316 for Og and Z=120 isotopes, respectively, due to a substantial chance of SF, it is very unlikely to detect any cluster or $\alpha$-decay by the experiments. The cluster decay half-lives for lower $A$ region are found smaller in comparison with $\alpha$-decay which endorses the significant probability of cluster decay over $\alpha$-decay. For the nuclei with 302$\leq$A$\leq$306 for Og isotopes and the nuclei with 312$\leq$A$\leq$316, a competition among considered decay modes is clearly evident, and hence in this region, a more detailed and separate investigation is required. It is also noticeable that in the considered region, the only experimental $\alpha$-decay half-live of $^{294}$Og matches excellently with the estimation of RRF for $\alpha$-decay and can be seen from the upper panel of Fig. \ref{half-lives-all}.\par

\begin{table}[!htbp]
\caption{Most probable cluster transitions in Og and Z=120 isotopes and their corresponding half-lives calculated by using RRF. $\alpha$-decay half-lives are also provided for a comparison.}
\centering
\def\arraystretch{0.8}
\resizebox{0.8\textwidth}{!}{%
\begin{tabular}{|c|cc|c|cc|}
\hline
\multicolumn{1}{|c|}{Transition}&
\multicolumn{2}{c|}{log$_{10}$T$_{1/2}$(Sec.)}&
\multicolumn{1}{c|}{Transition}&
\multicolumn{2}{c|}{log$_{10}$T$_{1/2}$(Sec.)}\\
\cline{2-3} \cline{5-6}
&Cluster &$\alpha$&&Cluster& $\alpha$\\
\hline
$^{284}$Og$\longrightarrow$$^{198}$Pb+$^{86}$Kr &      -12.99  &  -4.99  &  $^{290}$120$\longrightarrow$$^{202}$Pb+$^{88}$Se &     -16.48  &  -5.38 \\
$^{286}$Og$\longrightarrow$$^{200}$Pb+$^{86}$Kr &      -9.96   &  -4.43  &  $^{292}$120$\longrightarrow$$^{204}$Pb+$^{88}$Se &     -15.60  &  -4.99 \\
$^{288}$Og$\longrightarrow$$^{202}$Pb+$^{86}$Kr &      -10.67  &  -3.89  &  $^{294}$120$\longrightarrow$$^{206}$Pb+$^{88}$Se &     -15.16  &  -4.60 \\
$^{290}$Og$\longrightarrow$$^{204}$Pb+$^{86}$Kr &      -11.31  &  -3.90  &  $^{296}$120$\longrightarrow$$^{206}$Pb+$^{90}$Se &     -15.86  &  -4.84 \\
$^{292}$Og$\longrightarrow$$^{206}$Pb+$^{86}$Kr &      -12.09  &  -3.21  &  $^{298}$120$\longrightarrow$$^{208}$Pb+$^{90}$Se &     -15.17  &  -4.23 \\
$^{294}$Og$\longrightarrow$$^{206}$Pb+$^{88}$Kr &      -10.77  &  -3.17  &  $^{300}$120$\longrightarrow$$^{208}$Pb+$^{92}$Se &     -15.92  &  -4.88 \\
$^{296}$Og$\longrightarrow$$^{208}$Pb+$^{88}$Kr &      -11.27  &  -2.26  &  $^{302}$120$\longrightarrow$$^{208}$Pb+$^{94}$Se &     -15.61  &  -4.09 \\
$^{298}$Og$\longrightarrow$$^{208}$Pb+$^{90}$Kr &      -9.99   &  -3.21  &  $^{304}$120$\longrightarrow$$^{208}$Pb+$^{96}$Se &     -12.30  &  -3.88 \\
$^{300}$Og$\longrightarrow$$^{208}$Pb+$^{92}$Kr &      -6.58   &  -2.78  &  $^{306}$120$\longrightarrow$$^{208}$Pb+$^{98}$Se &     -12.90  &  -5.91 \\
$^{302}$Og$\longrightarrow$$^{208}$Pb+$^{94}$Kr &      -1.67   &  -2.99  &  $^{308}$120$\longrightarrow$$^{208}$Pb+$^{100}$Se &    -12.06  &  -4.36 \\
$^{304}$Og$\longrightarrow$$^{208}$Pb+$^{96}$Kr &      -0.82   &  -5.23  &  $^{310}$120$\longrightarrow$$^{210}$Pb+$^{100}$Se &    -9.52   &  -1.30 \\
$^{306}$Og$\longrightarrow$$^{210}$Pb+$^{96}$Kr &      -1.18   &  -3.99  &  $^{312}$120$\longrightarrow$$^{212}$Pb+$^{100}$Se &    -7.00   &  -0.69 \\
$^{284}$120$\longrightarrow$$^{196}$Pb+$^{88}$Se &     -17.78  &  -5.77  &  $^{314}$120$\longrightarrow$$^{214}$Pb+$^{100}$Se &    -4.17   &   0.39 \\
$^{286}$120$\longrightarrow$$^{198}$Pb+$^{88}$Se &     -17.53  &  -5.88  &  $^{316}$120$\longrightarrow$$^{214}$Pb+$^{102}$Se &      1.03   &  4.79 \\
$^{288}$120$\longrightarrow$$^{200}$Pb+$^{88}$Se &     -16.91  &  -5.38  &                     -                           &     -       &   -    \\   
\hline
\end{tabular}}
\label{cluster-half-live-result}
\end{table}
As a main result, in Table \ref{cluster-half-live-result}, we list the nuclei in which cluster decay is found dominant over any other decay modes for Og and Z=120 isotopes. In the Table, cluster decay half-lives estimated by RRF, and the corresponding $\alpha$-decay half-lives are mentioned for comparison. The listed clusters are expected to render additional impetus and motivation for the experimentalists and theoreticians eyeing to detect new elements with various possible decays, particularly in the superheavy region of the periodic chart.   

\section{Conclusions}
This article mainly focuses on the accurate estimation of half-lives of heavy cluster decay from superheavy nuclei where several widely used (semi)empirical/analytical formulas go wrong. We investigate this issue systematically and reach the conclusion that the formulas mainly based on the fission process work fairly well in this region of the periodic chart. The refitted version of the Royer formula (RRF) is found successful in producing cluster decay half-lives in superheavy nuclei estimated by various established theories. In addition, the same formula is fitted to 423 $\alpha$-decay data and found very precise in the estimation of half-lives. Hence, with the use of an asymmetric fission-like model for both cluster and $\alpha$-decay, a competition is probed which leads to favor several heavy cluster emissions from Og and Z=120 isotopes resulting in daughter nucleus as one of the isotopes of the closed shell nucleus Pb. The emission of magic daughter nuclei with Z$=$82 shows the impact of shell and pairing effects in the decay mechanism. 

Such accurate estimation in the unknown region is further embedded by the comparatively precise $Q$-values which are being tested from 18 theoretical approaches. The selection of the model for $Q$-values and the use of RRF in the superheavy region provides several Kr and Se isotopes as the most probable clusters from Og and Z=120 isotopes, demonstrating the dominance of the idea of heavy-particle radioactivity, which could show a new path on the possibility of clustering in superheavy nuclei and support the unexplored area for the synthesis of elements in the superheavy mass region in the future.

\section{Acknowledgement}
GS and MA acknowledge the support provided by SERB (DST), Govt. of India under SIR/2022/000566 and SR/WOS-A/PM-30/2019 scheme, respectively.

\end{document}